\documentclass[prb,showpacs,twocolumn,aps]{revtex4-1}
\usepackage{graphicx}
\usepackage{amsmath}
\usepackage{amssymb}
\usepackage{epsfig}
\usepackage{color}
\begin{document}


\title{Time dependent Faraday rotation}

\author{ Zhyrair Gevorkian$^{1,2,*}$,Vladimir Gasparian$^{3}$ and Josh Lofy $^{3}$},

\affiliation{$^{1}$ Institute of Radiophysics and Electronics,Ashtarak-2,0203,Armenia. \\
$^{2}$ Yerevan Physics Institute,Alikhanian Brothers St. 2,0036 Yerevan, Armenia.\\
$^{3}$ California State University, Bakersfield  California 93311-1022,USA}







\begin{abstract}
Faraday rotation in a magnetoactive medium with time dependent dielectric permittivity tensor is analyzed through both its diagonal and non-diagonal elements. Continuous and pulse incident laser field cases are considered. In a continuous case linear increasing of Faraday rotation angle with time is obtained.
In the continuous laser field case Faraday angle of rotation in both
time dependent diagonal and non-diagonal
element cases shows an increase with periodic oscillations as either positive (time-dependent dielectric  permittivity case) or negative (time dependent gyration vector case) and follows a general pattern.
Ultrashort pulse can scan the time dependent dielectric permittivity through the Faraday rotation angle. 
\end{abstract}


\maketitle

\section{Introduction}
In many situations a medium dielectric permittivity tensor  diagonal and non-diagonal elements vary with the time.
So far primary attention has been focused on systems with time dependent refractive index (see, for example Refs. \cite{xam:11,hayrapetyan:15} and references therein). However, in many situations the non-diagonal dielectric permittivity term ($g$) can also vary with time, originating time-dependent Faraday rotation (TDFR) \cite{Dali:85}. This justifies the study of both diagonal and non-diagonal elements for describing TDFR. Recently TDFR has been used, based on a left (right) circularly polarized pump-probe technique, for investigation of spin relaxation processes of a system using a continuous laser field \cite{PRL:12} or laser pulse \cite{nano:15} 
, or to study the effect of electron phonon coupling in metals undergoing thermal transport \cite{PRL:90}. In this technique a pump pulse is used to excite the sample, changing the optical properties while a second time-delayed probe pulse measures the change.

The main question that is addressed in this manuscript is whether or not the TDFR angle follows the time dependence of the dielectric permittivity non-diagonal or diagonal term. To answer this question it is useful to first establish a relation between Faraday rotation and the time dependent dielectric permittivity non-diagonal term. To the best of our knowledge, no such calculations have been previously reported.  The purpose of the present work goes in this direction in the sense that we provide analytical expressions for TDFR in the presence of pump-probe pulses and continuous laser field. Once the desired relation is obtained, one can analyze the rotational angle dependence on the characteristic parameters of the incident (probe) light. Finally, for completeness, we also consider TDFR dependence of the dielectric permittivity’s diagonal term. In this article we will restrict ourselves to a discussion of the slowly varying function of the time-dependent dielectric permittivity in a Gaussian pulse compared to the oscillations of a light-probe  in the case where $\Omega/\omega\ll 1$, with $\Omega$ being the frequency of oscillation of the gyration vector, and $\omega$ the frequency of light undergoing TDFR.
\section{Initial relations}
To start, let us look at Maxwell's equations without any external sources
\begin{equation}
rot{\bf E}=-\frac{1}{c}\frac{\partial{\bf B}}{\partial t},\quad rot {\bf H}=\frac{1}{c}\frac{\partial{\bf D}}{\partial t}
\label{maxeqs}
\end{equation}
where ${\bf E}$ and ${\bf B}$ are electric and magnetic fields, respectively. ${\bf D}$ is the electric displacement vector, and ${\bf H}$ is the magnetic displacement vector.  In a medium these equations should be completed with the material equations. For weak field case the connection between $\bf E$ and $\bf D$ remains linear
\begin{eqnarray}
D_i({\bf r},t)=\int_{-\infty}^t\varepsilon_{ij}(t,t-t^{\prime})E_j({\bf r},t^{\prime})dt^{\prime}\nonumber\\
E_i({\bf r},t)=\int_{-\infty}^t\varepsilon^{-1}_{ij}(t,t-t^{\prime})D_j({\bf r},t^{\prime})dt^{\prime}
\end{eqnarray}
where $\varepsilon_{ij}$ and $\varepsilon^{-1}_{ij}$ are the time-dependent dielectric and inverse dielectric permittivity tensors of the considered material.
Going to the Fourier transforms
\begin{equation}
\label{fourier}
D_i({\bf r},t)=\int\frac{d\omega}{2\pi}\varepsilon_{ij}(t,\omega)E_j({\bf r},\omega)e^{i\omega t}
\end{equation}
and assuming that for a non-dispersive medium $\varepsilon(t,\omega)\approx \varepsilon(t)$, one has $D_i({\bf r},t)\approx \varepsilon_{ij}(t)E_j({\bf r},t)$.
In an analogous manner $E_i(t)\approx \varepsilon^{-1}_{ij}(t)D_j(t)$.

We are going to consider situations when there are two time scales in the problem. The fast time scale is associated with the probe wave and the slow time scale is related to the external(magnetic, electric) fields. We take into account the slow time variable adiabatically assuming that the corresponding frequency is much smaller than the probe wave frequency.
Some specific physical realizations are presented below.

Using these relations and assuming that magnetic permeability $m_{ij}=\delta_{ij}$ for a non-magnetic medium we get
\begin{equation}
\Delta D_i({\bf r},t)-\frac{\varepsilon_{ij}(t)}{c^2}\frac{\partial^2D_j(t)}{\partial t^2}=0
\label{wave2}
\end{equation}
As for the dielectric permittivity tensor, it is assumed to have a form $\varepsilon_{ij}(t)=\varepsilon(t)\delta_{ij}+ie_{ijl}g_l(t)$,
where $e_{ijl}$ is the antisymmetric tensor and ${\bf g}$ is the gyration vector (see for example Ref. \cite{Landau:}).
  Suppose that the gyration vector  is directed in the wave propagation $z$ direction. Then the wave equation acquires the form
 \begin{equation}
 \Delta D_i({\bf r},t)-\frac{\varepsilon(t)}{c^2}\frac{\partial^2D_i({\bf r},t)}{\partial t^2}-ie_{ijz} \frac{g(t)}{c^2}\frac{\partial^2D_j({\bf r},t)}{\partial t^2}=0
 \label{wave2}
 \end{equation}
 which is valid for a continuous material with an arbitrary shape of the time-dependent dielectric permittivity.
Next, by introducing circular electric displacement components $D_{\pm}=D_x\pm iD_y$, the above equation becomes
 \begin{equation}
 \Delta D_{\pm}({\bf r},t)-\frac{\varepsilon_{\pm}(t)}{c^2}\frac{\partial^2D_{\pm}({\bf r},t)}{\partial t^2}=0
\label{separat}
\end{equation}
where $\varepsilon_{\pm}(t)=\varepsilon(t)\pm g(t)$.
Finally, assuming that the solution for complex $D_{\pm}$ is known, the time dependent Faraday rotation angle can be defined by
\begin{equation}\label{farangle}
  \tan\theta(L,t)=-i\frac{D_+(L,t)-D_-(L,t)}{D_+(L,t)+D_-(L,t)}
\end{equation}
where $L$ is the distance that light has propagated in the system.
\section{Rotation angle}
After these preliminaries we are ready for the evaluation of time-dependent Faraday angle $\theta(L,t)$, Eq. (\ref{farangle}), in several different cases.

The first case that should be examined is the static
one when $\varepsilon(t)=\varepsilon\equiv constant$ and $g(t)\equiv g =constant$. In this case, the solution of the wave equation (\ref{separat}) for electric displacement has the form of a monochromatic plane wave with the frequency $\omega$ propagating along the $z$-axis ($D^0$ is the initial magnitude of the electric displacement)
\begin{equation}\label{static}
D_{\pm}^0(z,t)=D^0e^{ik_{\pm}z-i\omega t},
\end{equation}
where $k_{\pm}={\omega\sqrt{\varepsilon\pm g}}/{c}$ represent forward propagating waves and $\pm$ indicates right-handed  and left-handed circularly polarized modes, respectively. Substituting Eq.(\ref{static}) into Eq.(\ref{farangle}), we are now able to reflect the case for bulk and isotropic materials where one-way light propagation is important in order to retrieve the standard expression for the angle of rotation of the polarization light.
\begin{equation}
\theta=\frac{L(k_+-k_-)}{2}=\frac{\omega L(\sqrt{\varepsilon+g}-\sqrt{\varepsilon-g})}{2c}.
\label{angle}
\end{equation}
Note that $\theta$ is positive for all normal (right-handed) materials.

The next case that we consider is the relevant dynamic case, where on top of the static case, discussed previously, the off-diagonal elements of dielectric permittivity tensor will have changed in time. This can be done
most straightforwardly by applying an {\it ac} external magnetic field (see, for example Refs. \cite{val:,jain:}) or use left (right) circularly polarized pump-probe technique without any externally applied magnetic field (see for example,  Ref. \cite{nano:15}). In the former cases the {\it ac} external magnetic field in the range of 20-40 G rms was used for determining Verdet constant.  In the latter cases \cite{nano:15,PRL:12} the polarized pump pulses were used for tracking the changes to the polarized carrier populations in time.

To simplify matters, we will consider a simplified
model of a pump
pulses or {\it ac} external magnetic field, assuming that
the time-dependent dielectric permittivity in a Gaussian
pulse is a slowly varying function of time compared to the oscillations of light-probe, i.e., $\Omega/\omega\ll 1.$

To proceed further, we seek solutions to the Maxwell's equations (\ref{separat}), which include the
slow variation in time of the $\varepsilon_{\pm}(t)$, in the form of a time-dependent Gaussian wave packet
\begin{equation}
D_{\pm}(z,t)=D^0e^{ik_{\pm}z-i\omega t-t^2/2T^2}F_{\pm}(\Omega t),
\label{anzats}
\end{equation}
where $T$ is the duration of pulse width, $\omega$ is the carrier frequency and $k_{\pm}=\frac{\omega}{c}\sqrt{\varepsilon_{\pm}(t=0)}$.
$F$ describes the influence of the slow varying portion of a laser pump pulse. We assume that $F(t=0)=1$ to be able to recover the Faraday angle in the absence of modulation, such as the case of Eq. (\ref{angle}) where the light propagates through a uniform medium with a constant dielectric permittivity.
Substituting above ansatz, Eq.(\ref{anzats}), into wave equation Eq.(\ref{separat}) and neglecting the second derivative of $F$ with respect to the time, one can easily solve the remaining first order differential equation. Ignoring the second derivative of $F$ is justified because here we are interested in slow modulation of the non-diagonal elements of the permittivity tensor compared to the oscillations of light ($\Omega/\omega\ll 1$).

The exact solution of the first order equation when integrated
from the initial time 0 to some instant of time $t$ leads to
\begin{eqnarray}\label{solution}
\ln F_{\pm}(\tau)=\int_{0}^{\tau}\frac{\omega^2\varepsilon_{\pm}(0)d\tau}{2\varepsilon_{\pm}(\tau)(i\omega\Omega+\tau/T^2)}+
\frac{i\omega\tau}{2\Omega}+\frac{\tau^2}{4\Omega^2T^2}-\\ \nonumber
-\frac{1}{4}\ln\bigg(1+\bigg(\frac{\tau}{\omega\Omega T^2}\bigg)^2\bigg)+
\frac{i}{2}\arctan\frac{\tau}{\omega\Omega T^2}\nonumber,
\end{eqnarray}
where $\tau=\Omega t$ is the dimensionless time variable.
It turns out, that only the contribution of the first term in Eq. (\ref{solution}) is relevant in time-dependent Faraday rotation (see the definition of Faraday angle, Eq. (\ref{farangle})). Hence, in our further discussion other terms will be ignored and we focus on the explanation behind the asymptotic behavior of the mentioned first term. However, it is clear, that the contributions of ignored terms become important if someone is interested in characterizing the dynamics of waves in a non-stationary complex dielectric permittivity medium (see, for example \cite{hayrapetyan:15}).
To proceed further, let us consider two limiting cases: (i) $T\to 0$ and (ii) $T\to \infty$. The first case corresponds to the ultrashort pulse and the second one to the continuous laser field. While the second case is by far the most important, we will also briefly discuss the pulses of very short duration, similar to that used in experiment \cite{nano:15}. It follows from Eq.(\ref{anzats}) that in the ultrashort impulse case only the times $t\to 0$ play significant role. Substituting in Eq.(\ref{solution}) $\varepsilon(\tau)\approx \varepsilon(0)$ we find that $F_{+}(\tau)=F_{-}(\tau)$ and hence, Faraday rotation angle is determined by the static case expression Eq.(\ref{angle}) where $\varepsilon$ must be replaced by $\varepsilon(t=0)$ and $g$ by $g(t=0)$.
If the pulse is centered not at $t=0$, but rather at $t=t_0$, then the rotation angle will be determined by the same static expression Eq.(\ref{angle}) where now $\varepsilon$ must be replaced by $\varepsilon(t=t_0)$ and $g$ by $g(t=t_0)$.
Now let us consider the continuous laser field case, that is $T\to\infty$. Using Eq.(\ref{solution}), one finds
\begin{equation}\label{efplusmin}
F_{\pm}(\tau)=\exp\left[-i\frac{\omega\varepsilon_{\pm}(0)}{2\Omega}\int_{0}^{\tau}
\frac{d\tau}{\varepsilon_{\pm}(\tau)}\right].
\end{equation}
Substituting Eqs. (\ref{efplusmin}) and (\ref{anzats}) into Eq. (\ref{farangle}), for the time dependent Faraday rotation angle we obtain
\begin{equation}
\theta(L,t)=\frac{\omega L}{2c}(\sqrt{\varepsilon+g}-\sqrt{\varepsilon-g})+\frac{\omega}{4\Omega}
\int_{0}^{\tau}d\tau\left[\frac{\varepsilon_-(0)}{\varepsilon_-(\tau)}-\frac{\varepsilon_+(0)}{\varepsilon_+(\tau)}\right].
\label{main}
\end{equation}
The first term in Eq.(\ref{main}) represents the static contribution and the second term represents time-dependent contribution to Faraday rotation. To proceed further one needs the explicit form of $\varepsilon_{\pm}(\tau)$.
We will consider three situations: two related to time-dependent non-diagonal terms of the dielectric permittivity tensor, and one related to the time-dependent diagonal terms of the dielectric permittivity tensor.

For the non-diagonal elements, one case is associated with spin-relaxation processes in different systems \cite{PRL:12, nano:15} $g=g_0e^{-t/\tau_r}$, where $\tau_r$ is some relaxation time and $t\gg \tau_r$. Second case is caused by the modulated low frequency external magnetic field $g=g_0\cos\Omega t$. In both cases the effect of a weak external magnetic field on the diagonal term is of order $\mathcal{O}(H^2)$ and hence can be neglected. Starting from the assumption that $g\ll \varepsilon$ and replacing $\Omega$ by $1/\tau_r$ in Eq. (\ref{main}), we may find an expression (without the static portion) for the time-dependent Faraday rotation angle for an arbitrary $g(t)$.
\begin{equation}
\theta_t(t)=\frac{\omega\tau_r}{2{\varepsilon}}\int_{0}^{\tau}d\tau g(\tau)-\frac{\omega {g_0} t}{2{\varepsilon}}.
\label{1a}
\end{equation}
It is easy to see that for large time scales the last term in Eq. (\ref{1a}) is dominated and the final sign of $\theta_t(t)$ is negative and varies almost linearly. Particularly, for a $g=g_0e^{-t/\tau_r}$ ($\tau_r$ is a relaxation time that depends on the details of the experiment) the above equation reads:
\begin{equation}\label{relax}
  \theta_t(t)=\frac{\omega g_0}{2\varepsilon}\left(\tau_r-\tau_r e^{-t/\tau_r}-t\right).
\end{equation}
Similar linearity of $\theta_t(t)$ versus time is observed in experiment \cite{PRL:12}, where the spin dynamics in EuO thin films based on the time-resolved Faraday rotation trace was investigated. As it was demonstrated in Ref.\cite{PRL:12}, the time trace of Faraday rotation includes two dynamic magnetization processes; one is an enhancement of magnetization ($\theta (t) > 0$) having two decay components, the other is a subsequent demagnetization at larger time delays $400 ps - 1.1 ns$ where $\theta (t) < 0$.

Now we implement an analogous procedure for the calculations of $\theta_t(t)$, Eq. (\ref{main}), assuming that $g=g_0\cos\Omega t$.
Using above mentioned equation and calculating the integrals \cite{ryzhik:65}, assuming $\varepsilon=const$, for the time-dependent part of Faraday rotation angle one finds the following expression in the given time interval $[-\pi/2+n\pi\le \Omega t \le \pi/2+n\pi]$, where $n=0,1,2,\cdots$,

\begin{equation}\label{final}
\theta_n(t)=\frac{\omega}{2\Omega\sqrt{\varepsilon^2-g_0^2}}
\bigg[\theta_0(t)-g_0\pi n\bigg] ,
\end{equation}
where
\begin{equation}
\theta_0(t)={\varepsilon}\arctan\frac{g_0\sin\Omega t}{\sqrt{\varepsilon^2-g_0^2}}-{g_0}\arctan\frac{\varepsilon \tan\Omega t}{\sqrt{\varepsilon^2-g_0^2}}.
\label{wher}
\end{equation}
\begin{figure}[htbp]
\centering
\fbox{\includegraphics[width=\linewidth]{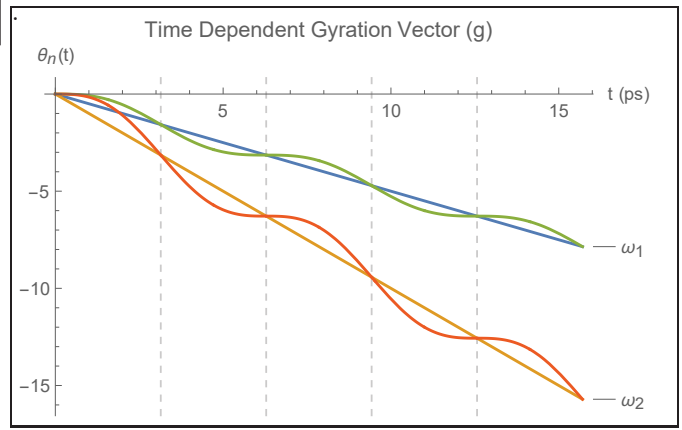}}
\caption{Schematic plot of time-dependent gyration vector Faraday rotation angle $\theta_n(t)$ (\ref{final}) vs. t. The parameters are: $g_0=0.01$, $\varepsilon=10$, $\omega_1= 10^{15}$, $\omega_2=2\times 10^{15}$ and $\Omega=10^{12}$.  The line $\frac{\omega g_0 t}{2 \sqrt{\epsilon^2-g_0^2}}$ is also plotted with $\theta_n(t)$ }
\label{fig:false-color}
\end{figure}

Eq. (\ref{final}) with Eq. (\ref{1a}) represents the central
results of this paper.

As shown in Fig. 1, the angle of Faraday rotation is constantly increasing (negatively) with a periodic change in slope which is illustrated by when the graph of $\theta_n(t)$ crosses the line $\frac{\omega g_0 t}{2 \sqrt{\epsilon^2-g^2}}$ every $\Omega t = n \pi$ with its oscillations following the general trend.  The gyration vector, $g$, is at its local maximum, $g_0$, every $\Omega t = 2 n \pi$ (where the derivative of $\theta_n$, $\frac{d \theta_n}{dt}$, becomes zero).  An increase of $\Omega$ will decrease period of oscillations, in contrast to increasing $\omega$ which will only increase the magnitude of amplitude of oscillations.  We show this schematically by choosing two different parameters of $\omega$ in all time dependent figures.  A similar periodic behavior has also been shown with Faraday rotation in films considering the length dependent case with multiple reflections  (see, e.g., Refs. \cite{Josh2016,proc}). In both the length dependent and time dependent gyration vector cases, the end result has increasing Faraday rotation angle with periodic oscillations as correspondingly positive (the length dependent case with reflections) or negative (the time dependent gyration vector case)

If the value of $g$ is restricted by the condition $g\ll\varepsilon$ (which is true in most materials), then $\theta_n(t)$ reads
\begin{equation}\label{simple}
\theta_{n=0}(t)=\frac{\omega g_0}{2\varepsilon}\bigg(\frac{\sin\Omega t}{\Omega}-t\bigg).
\end{equation}
As seen from Eq.(\ref{simple}), on the scale set by $\Omega t>1$, the main role is played by the second term that increases linearly with time $t$. It is opposite to the static terms sign (see Eq. (\ref{main})) and after some time can exceed it, thus changing the overall sign of the Faraday rotation angle.

To avoid misunderstanding note that the formulae Eqs.(\ref{simple}) and (\ref{relax}) are not applicable at very large times $t\gg 1/\Omega,\tau_r$. In this case the neglection of second derivative of $F(\tau)$ is not justified, see Eq.(\ref{solution}).

Also, if one considers the limit ${\varepsilon \rightarrow g_0}$, then the limiting value of $\theta_o(t)$
is
\begin{equation}
\lim_{\varepsilon \rightarrow g_0} \theta_0(t)=-\frac{\omega}{2\Omega}\tan\frac{\Omega t}{2}.
\end{equation}
The latter case can be realized in metamaterials , where, in order to reduce the diagonal elements of the permittivity tensor, one can introduce metal wires in the magneto-optical host medium (see, e.g., Ref. \cite{metam:16}).

\begin{figure}[htbp]
\centering
\fbox{\includegraphics[width=\linewidth]{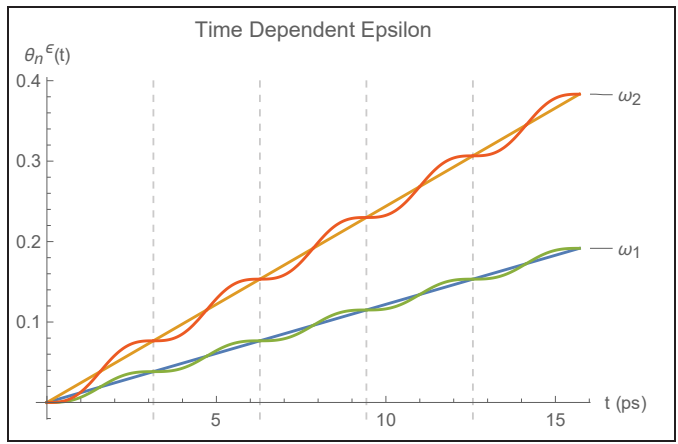}}
\caption{Schematic plot of time-dependent portion of Faraday rotation angle $\theta_n^\epsilon$ based on Eq. (\ref{epsilontime}).  The parameters are: $g_o=0.01$, $\epsilon_o=10$, $\epsilon_1=0.5$, $\omega_1= 10^{15}$, $\omega_2=2\times 10^{15}$, and $\Omega = 10^{12}$.  The line  $\frac{\omega}{4\Omega}(B-A)t$ is also plotted to schematically show the increase in Faraday rotation angle.}
\label{fig:false-color}
\end{figure}

Finally, for completeness, let us discuss the case where $\epsilon$ also depends on time.  To make the issue more clear we consider a simplified model following the case of Ref. \cite{hayrapetyan:15} where the time-dependent parity-time symmetric optical potentials are discussed, and assume that $\epsilon(t)=\epsilon_0+\epsilon_1 \cos^2(\Omega t)$ ($\epsilon_0$ is the non-time dependent dielectric properties of the system and $\epsilon_1$ is the amplitude of the variation with time). By using the same integral definition of Faraday angle, Eq. (\ref{main}), and calculating the integrals ($g$ is constant), we obtain the following result
\begin{eqnarray}\label{epsilontime}
\theta_n^\epsilon(t)=\frac{\omega}{4\Omega}
\bigg[A\arctan(A\cot(\Omega t))\\ \nonumber
-B\arctan(B\cot(\Omega t))+\pi\bigg(n+\frac{1}{2}\bigg)(B -A )\bigg]
\end{eqnarray}
where $A=\frac{\sqrt{{\epsilon_0}+\epsilon_1+g}}{\sqrt{{\epsilon_0}+g}}$,
$B=\frac{\sqrt{{\epsilon_0}+\epsilon_1-g}}{\sqrt{{\epsilon_0}-g}}$, and $n=0,1,2,\cdots$.
The time interval for which the expression (\ref{epsilontime}) is valid $[ n\pi \le \Omega t \le (1+n)\pi ]$.

The resulting graph is plotted in Fig. 2.
This result is very similar to the time dependent case of the gyration vector increasing along the line $(B-A)t$ and crossing the line every  $\Omega t = \frac{n\pi}{2}$, with the primary difference being that the increasing angle is always positive rather than negative. In both cases the time dependent diagonal or non-diagonal elements have a line  that the Faraday rotation angle oscillates about and crosses multiple times, attributing a periodic and constantly increasing rotation angle as time passes.
\section{Summary}
What we have shown, with the assumption of a slow modulation of the non-diagonal elements of the permittivity tensor when compared to the oscillations of light within a system where ($\Omega/\omega\ll 1.$) and similar to that of perovskite semiconductor thin films \cite{nano:15}, is that it is possible to calculate the time-dependent Faraday angle $\theta(L,t)$, Eq. (\ref{farangle}), for two different cases using the time-dependent non-diagonal and diagonal terms of the dielectric permittivity tensor. One case is caused by the modulated low frequency external magnetic field $H$ that leads to the slow time-dependent variation of the gyration vector $g=g_0\cos\Omega t$. The second case is associated with ultrafast magneto-optical experiments in ferromagnets where the dynamics of the electrons spin degree of freedom were investigated (see, e.g., Ref. \cite{xam:11} and references therein). In the latter case for the gyration vector we took the form, $g=g_0e^{-t/\tau_r}$ ($\tau_r$ is some relaxation time). In both cases, the time trace of $\theta (t)$ clearly indicates a change in sign as measured in Ref. \cite{PRL:12}.

 We have also calculated the time dependence of the diagonal term of the dielectric tensor on the Faraday rotation angle, Eq. (\ref{epsilontime}), where the diagonal element $\epsilon$ is defined as $\epsilon(t)=\epsilon_o+\epsilon_1 \cos^2 \Omega t$ and $g=const$.  These calculations of the Faraday rotation angle are varying between $0$ and $2\pi$, but the representation that we show represents a constant increase in magnitude, positive for the diagonal element and negative for the non-diagonal element.  This constant increase represents a continuous laser field’s effects.




\section{Acknowledgments}
Zh.G. is grateful to A.Hakoumian for pointing out his attention to this problem and for helpful discussions, V.G. acknowledges partial support by Fundaci\'on S\' eneca grant   No. 19907/GERM/15.

\end{document}